# Quantitative Analysis of Narrative Reports of Psychedelic Drugs

Jeremy R. Coyle[1]
David E. Presti[2]
Matthew J. Baggott[3]§

[1]Graduate Group in Biostatistics, University of California, Berkeley, California 94720-7358 USA
[2]University of California, Department of Molecular and Cell Biology, Berkeley, California 94720-3200 USA
[3]Human Behavioral Pharmacology Laboratory, Department of Psychiatry and Behavioral Neuroscience, University of Chicago, Chicago, Illinois 60637 USA
§Tel: (773) 702-0563, Email: matthew@baggott.net

**ABSTRACT**

**BACKGROUND:** Psychedelic drugs facilitate profound changes in consciousness and have potential to provide insights into the nature of human mental processes and their relation to brain physiology. Yet published scientific literature reflects a very limited understanding of the effects of these drugs, especially for newer synthetic compounds. The number of clinical trials and range of drugs formally studied is dwarfed by the number of written descriptions of the many drugs taken by people. Analysis of these descriptions using machine-learning techniques can provide a framework for learning about these drug use experiences.
**METHODS:** We collected 1000 reports of 10 drugs from the drug information website Erowid.org and formed a term-document frequency matrix. Using variable selection and a random-forest classifier, we identified a subset of words that differentiated between drugs.
**RESULTS:** A random forest using a subset of 110 predictor variables classified with accuracy comparable to a random forest using the full set of 3934 predictors. Our estimated accuracy was 51.1%, which compares favorably to the 10% expected from chance. Reports of MDMA had the highest accuracy at 86.9%; those describing DPT had the lowest at 20.1%. Hierarchical clustering suggested similarities between certain drugs, such as DMT and *Salvia divinorum*.
**CONCLUSION:** Machine-learning techniques can reveal consistencies in descriptions of drug use experiences that vary by drug class. This may be useful for developing hypotheses about the pharmacology and toxicity of new and poorly characterized drugs.

## INTRODUCTION

Psychedelic substances such as lysergic acid diethylamide (LSD) and 3,4-methylenedioxymethamphetamine (MDMA; street name "ecstasy") are widely consumed. An estimated 14.8% of Americans over the age of 12 years have used psychedelics at some point, commonly LSD (9.2% of the population) and ecstasy (6.3%) [1]. In Europe, psychedelic use varies greatly by country: across 22 countries, the by-country median percent of the adult population (ages 15–64 years) having used LSD and ecstasy is estimated at 1.3% and 2.3%, respectively [2]. Despite widespread use, these drugs are poorly understood, largely because formal controlled experiments in humans are rare. In this report, we use a machine-learning approach to understanding descriptions of psychedelic experiences.

Psychedelics facilitate a variety of complex mental states, including alterations in perception, intensification of emotions and thoughts, and changed sense of boundary between the self and environment [3, 4]. They have been used experimentally to imitate psychosis [5-7] yet they have also been reported to facilitate mystical and spiritual experiences [8, 9]. The complexity of psychedelics' effects is underscored by the many terms used to label them: psychedelics; hallucinogens; psychotomimetics; and entheogens. Because of these complex effects, psychedelics are powerful probes of the connection between brain physiology and consciousness, one of the most mysterious problems in neuroscience.

Psychedelic use by humans may predate written history. In their plant and fungal forms, psychedelics have probably been used for millennia for therapeutic, ritualistic, and religious purposes [10, 11]. Modern scientific research on psychedelics can be traced back to Arthur Heffter's [12] isolation of mescaline from the peyote cactus in the 1890s. The pace of scientific research quickened after Albert Hofmann's discovery of LSD in the 1940s. Research on LSD contributed to the recognition that serotonin functions as a brain neurotransmitter intimately involved in the regulation of mental processes [13]. In the 1950s and 1960s, psychedelics were extensively studied for their psychotherapeutic potential [14, 15]. This has continued in a limited manner [16, 17]. However, the majority of scientific publications on psychedelics use them to elucidate neuropharmacological signaling mechanisms in rodent subjects [18, 19].

These studies support a central role for the $5HT_{2A}$ receptor in the mechanisms of what are sometimes called serotonergic or classical psychedelics [3]. These classical compounds include LSD, N,N-dimethyltryptamine (DMT), mescaline, and psilocybin. However, many other compounds with different mechanisms of psychoactivity share clinical effects and users demographics with classical psychedelics. Examples of such chemicals are the monoamine releaser MDMA (methylenedioxymethamphetamine: street name "ecstasy"), the kappa-opioid receptor agonist salvinorin A (from the plant *Salvia divinorum*), and the NMDA-receptor antagonist anesthetic ketamine. The clinical similarities of these drugs suggests convergent downstream effects on consciousness despite neurochemically distinct initial mechanisms [6, 20]. Thus, documenting the clinical and experiential effects of different psychedelics can elucidate connections between brain physiology and consciousness.

Studying psychedelics also has potential practical value. New psychedelics — many first synthesized as probes of the brain-mind connection [21-23] — have become available in the unregulated market in recent decades [24]. These drugs are often sold as "research chemicals" or "legal highs," but are also sometimes disingenuously marketed as "bath salts", "herbal incense", or "plant food" [25, 26]. This disingenuous marketing is common in the US, because the Federal Controlled Substance Analogue Act bans many novel hallucinogens only when they are intended for human consumption. Use of novel compounds has led to adverse reactions, especially in situations where unknown doses of unknown compounds are involved [25-27]. Understanding the human pharmacology of these compounds could aid in reducing and treating their toxicities. However, the scientific literature is virtually silent about the effects of most of these novel chemicals in humans in controlled settings. The expense associated with doing human research ensures this situation will change only very slowly.

In contrast to the limited formal scientific literature, there is a robust informal literature available on the Web written by individuals who use psychoactives. An important source of such information is the website Erowid.org. This website receives approximately 95,000 unique visitors per day and contains narrative descriptions of nearly 22,000 drug-use episodes [28]. These drug-experience reports range from descriptions of caffeine, alcohol, and nicotine, to marijuana, LSD, and mescaline, to lesser-known substances such as 4-bromo-2,5-dimethoxyphenethylamine (2C-B) and 2,5-dimethoxy-4-ethylthiophenethylamine (2C-T-2). In some cases, these descriptions represent the only written sources of information about the effects of a specific compound in humans.

Given these considerations, we sought to develop a method for quantitative comparison of drug use descriptions archived online. Previous studies of written descriptions of drug effects have generally used either qualitative methods or dictionary approaches. Qualitative methods are time consuming and usually involve very limited sample sizes [29-31]. Dictionary approaches in which descriptions are scored based on the appearance of words from previously validated categories can be applied to larger data sets [32]; however, such approaches may fail to capture the changing terminology used to describe drug experiences. Machine-learning techniques such as classification and variable selection avoid some of the limitations of these approaches, as they can accommodate a large sample size and make classifications based on characteristics of the data set, rather than on predetermined categories.

We used classification techniques in an attempt to characterize differences in narrative reports from 10 psychedelic drugs. We hypothesized that:
   i) there would be detectable differences between reports for different drugs as indicated by the ability of a classifier to accurately predict drug class;
   ii) drugs with similar effects would have similar reports as indicated by classifier confusion and class means;
   iii) inspection of the discriminating variables would allow insight into the differences between these drugs.

We tested these hypotheses using pre-existing, publically available drug-experience reports from the website Erowid.org. We were able to successfully classify reports from 10 different drugs with an estimated overall accuracy of 51.1% and we furthermore identified a subset of 110 variables that were particularly useful in distinguishing drug classes. We conclude that machine-learning techniques provide a promising approach for gaining insights into the pharmacology, and toxicity of new and poorly characterized drugs.

**METHODS**

*Data Collection*
We collected a corpus of 1000 narrative reports from the drug-experience archives at Erowid.org. Narratives represented 10 psychedelically-active substances (Table 1), which were selected because their pharmacology is at least partly understood. Reports were written between 2000 and 2010. For each substance, we randomly chose 100 reports.

**Table 1:** Names, routes of administration, and pharmacology for the 10 studied drugs

| Chemical/Species Name | Street Names | Routes of administration | Primary Known Pharmacological Mechanisms |
|---|---|---|---|
| 2C-E (2,5-dimethoxy-4-ethylphenethylamine) | Europa | oral, insufflated | 5-HT$_{2A}$ agonist [33, 34] |
| 2C-T-2 (2,5-dimethoxy-4-ethylthio-phenethylamine) | T2 | oral, insufflated | 5-HT$_{2A}$ agonist [34] |
| 5-MeO-DiPT (N,N-diisopropyl-5-methoxytryptamine) | foxy methoxy, foxy | oral, insufflated, smoked | 5-HT$_{2A}$ agonist, binds to 5-HT$_{1A}$ receptors [33-36] |
| 5-MeO-DMT (5-methoxy-N,N-dimethyltryptamine) | 5-MeO | smoked, insufflated | 5-HT$_{1A}$ and $_{2A}$ agonist [33, 34, 36-39] |
| DMT (N,N-dimethyltryptamine) | Dimitri | oral, insufflated, smoked | 5-HT$_{2A}$ agonist; sigma-1 regulator [34, 40, 41]; trace amine receptor agonist [42] |
| DPT (N,N-dipropyltryptamine) | the light | oral, insufflated, smoked, intramuscular | 5-HT$_{1A}$ and $_{2A}$ agonist [34, 35]; SERT inhibitor/substrate [33, 43] |
| LSD (lysergic acid diethylamide) | acid, blotter, tabs | oral | 5-HT$_{2A}$ and $_{1A}$ agonist and D2-like agonist [44, 45] |
| MDMA (3,4-methylene-dioxymethamphetamine) | ecstasy, e, rolls, molly | oral, insufflated | Monoamine releaser [46] |
| Psilocybin Mushrooms (psilocin is active metabolite) | shrooms, magic mushrooms | oral | 5-HT$_{2A}$ and $_{1A}$ agonist [44, 47] |
| *Salvia divinorum* (Salvinorin A is active ingredient) | salvia, diviner's sage | smoked | Kappa opioid agonist [48] |

*Approach*

In order to identify the differences between the drugs in our analysis, we conducted a classification analysis in which frequencies of individual words in each report were used to predict which drug was reportedly taken. Classification techniques allowed us to quantify how difficult it is to differentiate between drugs (how similar they are), and which words were most useful for prediction of the drug being described. We used variable selection techniques to determine which of the large number of words were useful for accurate report classification. Finally, we clustered selected words based on their co-occurrence in reports to visualize and form qualitative descriptions of the differences between drugs and between groups of drugs.

*Preprocessing*

First, we checked reports for spelling errors in a word processor. We then converted each report to a vector with one element for each unique word, containing the number of occurrences of that word in the report. To reduce the dataset to a more manageable size, we dropped words that occurred in fewer than five reports. We converted all words to their root form with the Morphy stemmer [49], allowing all forms of a word to be treated as a single element. Morphy determines the root of a word using a dictionary. Words that were not present in the dictionary were stemmed manually. We then eliminated two types of words from the remainder of the analysis: common words [such as pronouns (e.g., 'he') and determiners (e.g., 'the')] and drug terminology. We removed common words because we considered them unlikely to differentiate between documents or drugs. We removed drug terminology (e.g., 'ecstasy' and 'roll') to focus the analysis on differences in drug effects instead of trivial differences. The document vectors were then combined into a term-document frequency matrix with a row for each report and a column for each unique word in the corpus. To control for differences in report lengths, we converted the word occurrence counts into word frequencies.

*Classification*

We applied Leo Breiman's random-forest classification algorithm [50]. Random forest is an ensemble classifier that generates a group of classification trees based on predictor variables and then uses the majority vote of the trees to determine membership. Each classification tree is fit using a random subset of predictor variables on a random subset of the observations drawn with replacement. The performance of this classifier can be assessed using out-of-bag accuracy by predicting each observation using only those trees for which the observation was out-of-bag, meaning that the observation was not used in the tree's construction. Variables can be selected on the basis of an importance measure, defined as the decrease in out-of-bag accuracy caused by randomly permuting the values of that predictor. Variable selection using random forests are commonly employed in microarray genetics experiments to select from a large pool of candidate genes those that predict a condition [51, 52].

*Feature Selection*

We used the varSelRF package in R [53, 54] for feature selection. This package uses a two-step algorithm to determine a minimal subset of predictor variables that classifies almost as well as all available predictors. First, a random forest is fit using all predictor words and these words are ranked using the importance measure described above. Then, new random forest models are fit to progressively smaller subsets of variables, and the performance of these reduced models is assessed using out-of-bag accuracy, as described above. The subsets are generated by eliminating the least important 20% of variables at each step. After all subsets are fit, varSelRF selected the smallest subset which classifies with an accuracy less than 1 standard error unit lower than the subset with the highest accuracy. This smaller subset, while not being significantly worse at differentiating the classes, allows for easier interpretation.

*Bootstrapping*
To measure classifier performance without overfitting, we used .632 bootstrapping. Before bootstrapping, we applied varSelRF to the original dataset. Then, at each of 200 iterations, we sampled the data with replacement to form a training set, and reserved the remaining data for testing. With the training set, we used varSelRF to select a set of predictor variables. We then fit a random forest model on the training set using the selected variables and assessing classifier accuracy using the testing set.

*Performance Assessment*
Different measures of classifier accuracy are used for different purposes. Most of these estimates are biased because resampling can lead to overfitting. Such estimates are valid only for determining relative performance under different conditions. We used out-of-bag accuracy (described above), to compare performance at different subset sizes during feature selection. Applying out-of-bag accuracy to the bootstrap iterations would have been problematic because reports are sampled with replacement, and therefore a duplicate version of a report could mean the trees are not truly out-of-bag. To determine accuracy for each drug and create a confusion matrix between drugs, we used accuracy of prediction for the testing data from each bootstrap iteration. For an unbiased estimate of overall performance we used the .632 bootstrap estimator, which combines accuracy for bootstrap testing data with the out-of-bag accuracy using the subset of variables selected from the original dataset.

*Stability*
We assessed the stability of the variable selection procedure by comparing the words selected from the original dataset to those selected in each of the bootstrap iterations. We then calculated the proportion of iterations in which each of these words appeared and used the median as an overall measure of stability [52, 54].

*Visualization*
To facilitate qualitative analysis of differences between drugs, we created a heatmap of mean word frequencies by drug for the words selected by varSelRF from the original dataset. We used hierarchical clustering analysis to form both word and drug clusters and depicted the clusters using dendrograms. Word clusters were formed using all the reports and drug clusters were formed using mean reports for each drug class.

*Software*
All analyses were conducted using R 2.11 [55] and the Revolution R development environment. Amazon's Elastic Compute Cloud was used for the computationally-intensive resampling procedures.

**RESULTS**

We obtained a total of 1000 reports with 22440 unique words. Of these, we removed 15994 words that occurred in fewer than five reports. Stemming the remaining 6445 words resulted in 4541 unique stems. We removed 607 of these that were common words or drug terminology, leaving a final count of 3934 unique word stems. The initial run of the variable selection procedure selected 110 variables. The subset of 110 variables had an out-of-bag accuracy of 46.2%, similar to the out-of-bag accuracy of the full data set (45.5%). Bootstrapping iterations selected a range of possible subset sizes from 29 variables to all 3934 variables, with a median size of 138 (Figure 1).

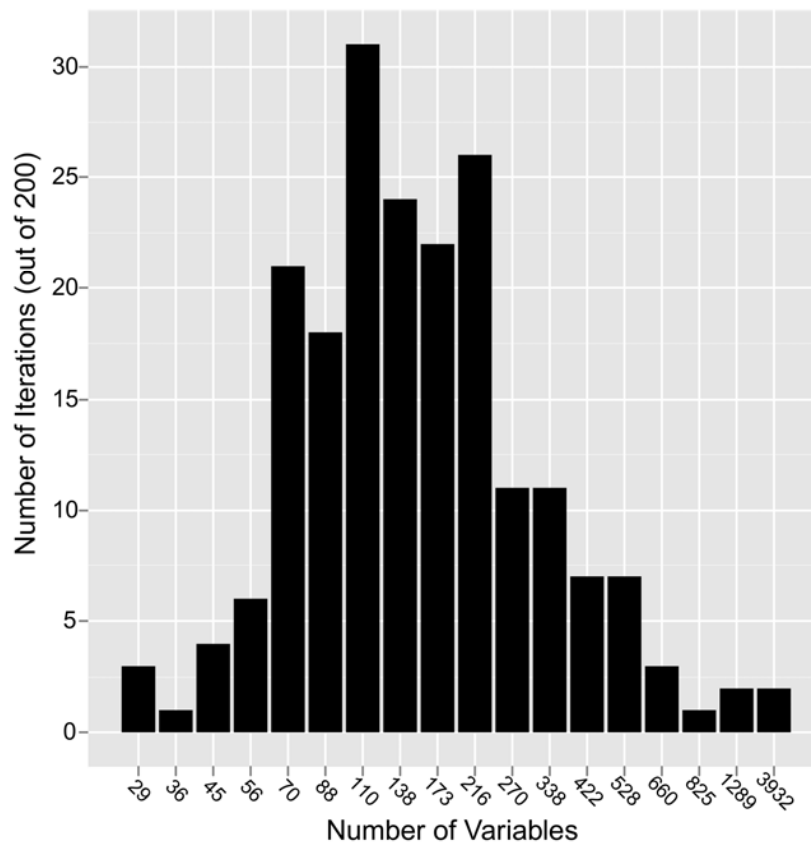

**Figure 1 – Distribution of subset sizes:** The number of predictor words selected by each of the 200 bootstrapping iterations. In each iteration, the algorithm selected a number of words that balanced parsimony with accuracy. In most iterations, only a few hundred words were needed to classify the reports with reasonable accuracy.

The .632 estimator determined that the classifier was 51.1% accurate. Individual class accuracy for bootstrap testing ranged from 86.9% for MDMA to 20.7% for DPT (Figure 2A). The classifier confusion matrix shown in Figure 2B indicates which drugs were mistaken for other drugs by the classifier. The subset of selected words was relatively stable. Words selected from original dataset were selected in the majority of bootstrap iterations (median proportion = 0.768, 1st quartile = 0.532, 3rd quartile = 0.984). Hierarchical clustering of class predictions and class means are depicted in Figures 2B and 3 (included as an Appendix). Figure 3 also depicts clusters of predictor words from the subset of 110 selected words.

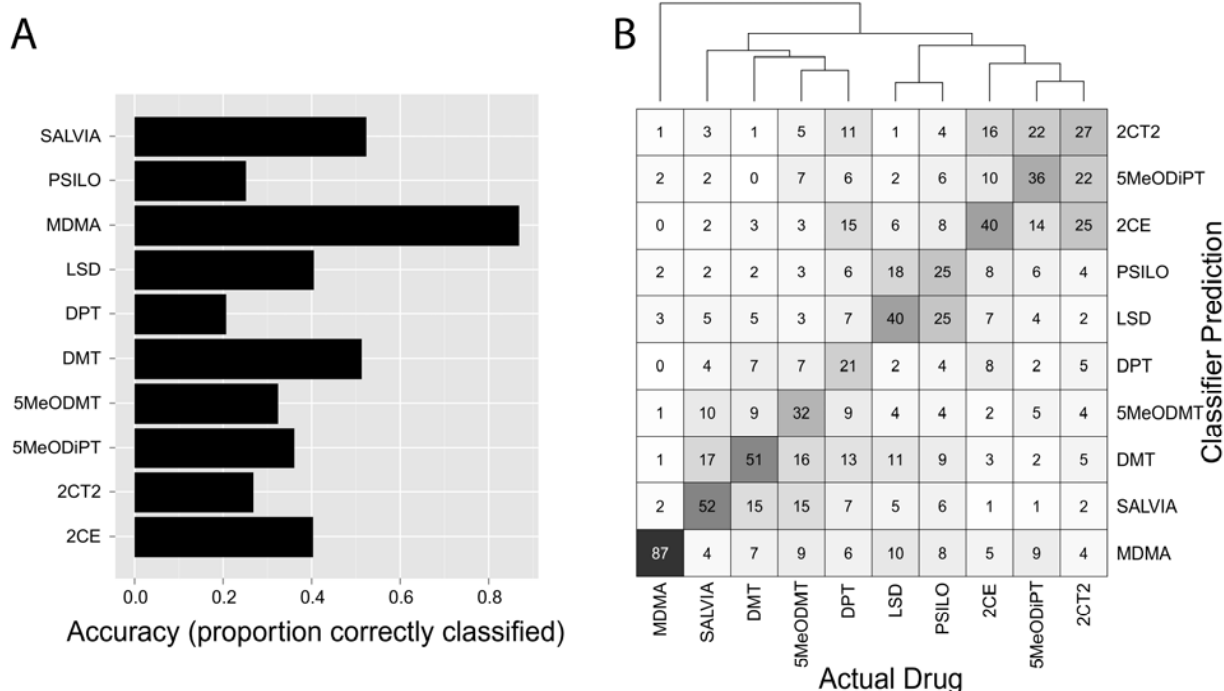

**Figure 2 – Classification accuracy and confusion matrix: (A)** Bootstrap accuracy (percent of left out training samples accuracy classified) for each of the 10 drugs. There is a wide range of accuracies. **(B)** Each cell represents the percent of reports from a given drug (in columns) predicted as being from another given drug (in rows). The dendrogram at the top clusters together drugs that are misclassified as each other, as this misclassification is related to similarities in reports from these drugs

**DISCUSSION**

Our success at classifying reports suggests that there are consistencies within narratives for a given drug and differences between narratives for different drugs. We used recursive feature selection to identify some of these consistencies and differences. This revealed that a small subset of words that was able to classify reports almost as well as the full set of 3934 words. The accuracy of classifying reports with these 110 words was estimated as 51.1%. This can be favorably compared to 10%, which would be expected from chance. The small subset of words allows a practical analysis of the differences between drugs. Specifically, our results suggest that narratives of MDMA use are more consistent than other psychedelics and that specific pairs of other drugs may share interesting features: for example, DMT and *Salvia*, LSD and psilocybin mushrooms, DPT and 2C-E, and 5-MeO-DiPT and 2C-T2 (Figure 2B).

To analyze the 110 words, we created a heat map and dendrograms of class centroid clustering of word frequencies and drugs. This was necessary because it would not have been practical to depict the many binary branching classifier trees in which these individual words appeared. Moreover, the selected words are not necessarily high-frequency words and are not easily summarized by frequency. Instead, they are words whose frequencies are useful for deciding between two potential drugs in the context of decision tree classifiers. The visualization of the 110 words (Figure 3, see Appendix) therefore provides different information from the confusion matrix of classifier performance (Figure 3). For example, the confusion matrix indicates that 5-MeO-DMT was often confused for DMT and *Salvia*, while the class centroid clustering visualization groups it with DPT and 2CE. This indicates that some of the differences between these drugs is captured in those words that were difficult to interpret and therefore not analyzed with class centroids (Figure 3).

MDMA was distinguished easily from all other drugs (estimated accuracy 86.9% vs 52.4% for *Salvia*, the next most accurate drug). This grouping of LSD and psilocybin is consistent with the hypothesis that MDMA is a member of a novel pharmacological class (Nichols 1986). MDMA reports were distinguished from other drugs by high frequencies of words with social connotations (e.g., the "club", "hug", "rub", "smile" cluster in Figure 3). Examination of individual reports suggests that this reflects descriptions of the social context of MDMA use (e.g., night clubs) as well as drug effects proper (e.g., feelings of love and friendliness [56]).

LSD and psilocybin mushrooms were also grouped together. These drugs (and to a lesser extent DMT and *Salvia*) were most strongly associated with the "see", "look", "saw", "room", "tell", "ask", "walk", "house" cluster. This grouping is consistent with early clinical comparisons finding that they produce similar effects in humans [57]. Alternatively, this clustering could be a product of these substances' longstanding availability compared to other psychedelics. This could lead to narrative differences between these reports and those of more exotic drugs since authors may have assumed reader familiarity with the drugs and been less focused on describing common symptoms.

DMT and *Salvia* were grouped together, with high frequency word clusters suggesting powerful effects on the perception of reality. The highest frequency cluster for these drugs was the "reality", "dimension", "universe", "state", "consciousness", "form", "entity" cluster. Both drugs have been associated with powerful alterations in consciousness and feelings of altered reality. Rick Strassman [58], summarizing his studies of administering DMT to healthy human volunteers, described higher dose hallucinatory experiences that were "more compelling and convincing than 'ordinary' reality or dreams." Similarly, in a controlled study of *Salvia*'s primary active chemical constituent, salvinorin A, participants reported intense experiences characterized by disruptions in vestibular and interoceptive signals (e.g., change in spatial orientation, pressure on the body) [59]. While DMT and *Salvia* have little overlap in their known pharmacological mechanisms (DMT is a mixed serotonergic agonist and salvinorin A is a kappa-opioid agonist), they are both typically taken by smoking or vaporizing.  Thus, the apparently similar intense drug effects may be explained by their common routes of administration and the rapid rises in blood concentrations of both drugs.

DPT and 2C-E were clustered together and were primarily associated with a "stomach", "nausea", "vomit", "headache" cluster, suggesting these drugs may have unpleasant physical effects.  Because $5\text{-}HT_3$ receptor stimulation is thought to induce nausea and vomiting [60], we hypothesize that these two substances, more than the other studied psychedelics, may directly stimulate $5\text{-}HT_3$ receptors or may induce release of 5-HT from enterochromaffin cells.  Both drugs are currently poorly characterized, although it was reported that DPT inhibits [(3)H]5-HT transport at the serotonin transporter with a $K_i$ of 0.594 μM [43].

*Limitations*
We assume that most individuals actually consumed the drugs that they reported and are writing in good faith.  Nonetheless, we cannot know how many drug reports are mislabeled or falsified in the current data set.  Notably, however, random forests are resistant to the effects of a few mislabeled cases (Breiman 2001).  Consistent with this expectation, our classifier was highly successful with MDMA, despite the well-documented sale of other drugs as MDMA [61].  Due to their distinctive appearances, it may be reasonable to assume that illicitly obtained psilocybin mushrooms, LSD, and *Salvia* were likely to be correctly represented.  The six other drugs, being sold as "research chemicals," may be more prone to misrepresentation in the unregulated marketing of crystalline or powdered chemicals.  Thus, it is noteworthy that one of the few available analyses of similar synthetic "research chemicals" purchased in the unregulated marketplace, were found them to be correctly represented as to chemical identity and also of relatively high purity [33].

Attempts to infer drug effects from narratives have several other inherent limitations.  Differences between drug narratives likely also reflect author demographics and the varying context of drug use.  Such potential limitations could be addressed by collecting a new corpus in which author demographic information is also obtained.  Additionally, from one perspective, the mixing of effects from uncertain drugs with context is as much a "feature" as a "bug" since such reports (if written in good faith) reflect drug experiences as they are understood by users, which is an important topic of study.

Further research on drug narratives might profitably categorize individual narratives based on route of administration and physical context in order to better understand the influence of these variables.  Most excitingly, perhaps, our approach could be used to link narrative descriptions to pharmacology.  Data on the *in vitro* pharmacology of many of the chemicals addressed in this paper is beginning to be assembled [34]. Quantitative analysis of drug narratives in combination with *in vitro* pharmacology could lead to novel hypotheses concerning the effects of specific receptors and signaling pathways on consciousness.

In conclusion, we present results from a novel approach to classifying and analyzing drug narratives. We find that drug narratives can be successfully classified using random-forest techniques, suggesting that this can be used to develop hypotheses about the pharmacology and toxicity of new and poorly characterized drugs as they may emerge.

**ACKNOWLEDGEMENTS**

We are grateful to Spoon, Earth, and Fire Erowid for their assistance in collecting and making available the narrative reports and for helpful discussion and inspiration. We thank Juan Carlos Lopez for programming assistance, and former UC Berkeley students Conor Penfold and Andrea Payne for assistance in early stages of this work.

**APPENDIX**

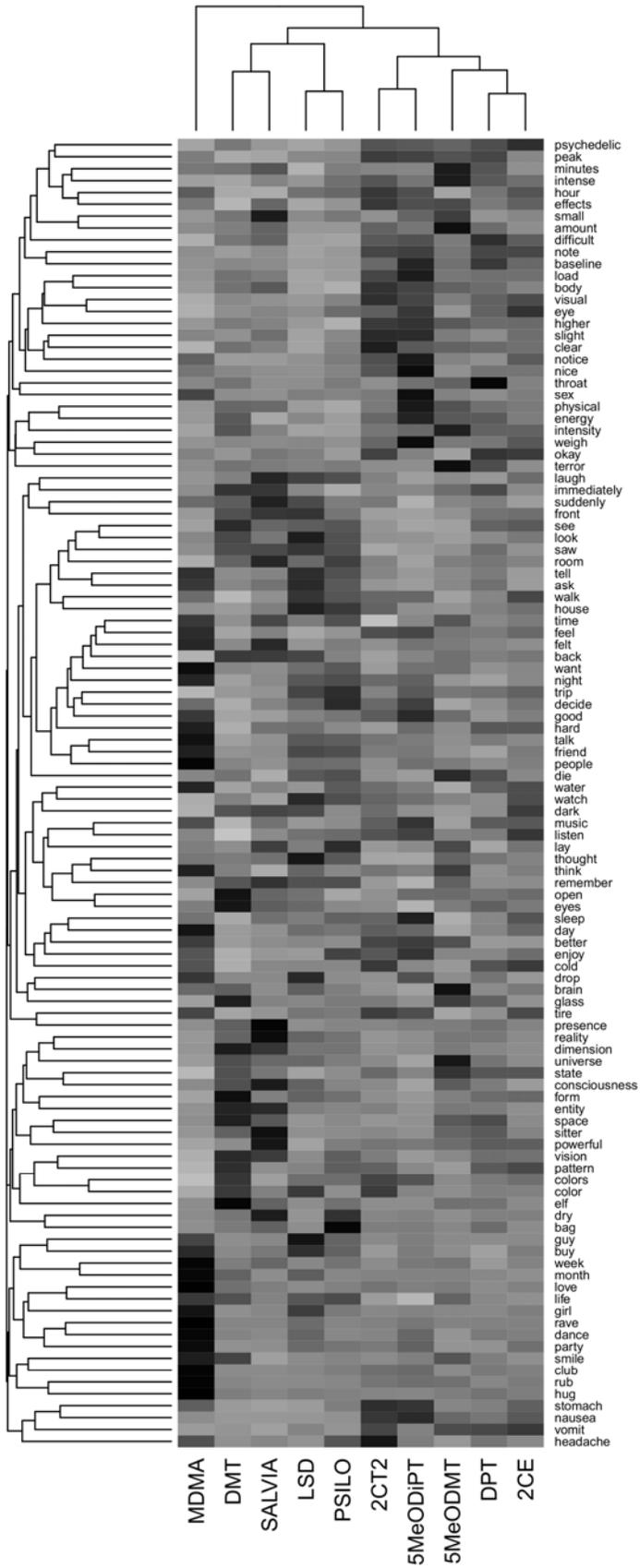

**Figure 3 – Class mean heatmap:** Each cell is the mean frequency of a given word (columns) in reports from a given drug (rows) in all reports from the data set. Darker cells represent higher frequencies. These words were selected as important for differentiating drugs by the random forest algorithm. Words are clustered based on co-occurrence in reports. Drugs are clustered based on similar mean frequencies for these words. These clusterings are represented by dendrograms at the right and top sides of the plot.